\documentclass[12pt]{article}
\usepackage{amssymb,amsmath}
\begin{document}
 \vspace{-7.2mm}
\title{\textbf{Nonlocal Gravitational Models\\ and Exact Solutions}}

\author{Sergey Yu. Vernov\footnote{E-mail: vernov@ieec.uab.es,
svernov@theory.sinp.msu.ru}\\
 \small  Instituto de Ciencias del Espacio,
 Institut d'Estudis Espacials de Catalunya  \\
\small  Campus UAB, Facultat de Ci\`encies, Torre C5-Parell-2a planta, \\
\small  E-08193, Bellaterra (Barcelona), Spain\\
\small  Skobeltsyn Institute of Nuclear Physics,  Lomonosov Moscow State University,\\
\small  Leninskie Gory 1, 119991, Moscow, Russia}
\date{ }
\maketitle
\vspace{-7.2mm}
\begin{abstract}
A nonlocal gravity model with a function $f(\Box^{-1} R)$, where $\Box$ is the d'Alembert ope\-rator,  is considered.
The algorithm, allowing to reconstruct $f(\Box^{-1} R)$, corresponding to the given Hubble parameter and
the state parameter of the matter, is proposed. Using this algorithm, we find the functions $f(\Box^{-1} R)$,
corresponding to de Sitter solutions.
\end{abstract}

\section{Nonlocal gravitational models}
In this paper we consider nonlocal
gravity models, which are describing by the action
\begin{equation}
\label{nl1} S=\int d^4 x \sqrt{-g}\left\{ \frac{1}{2\kappa^2}\left[
R\Bigl(1 + f\left(\Box^{-1}R \right)\Bigr) -2 \Lambda \right] +
\mathcal{L}_\mathrm{matter} \right\} \, ,
\end{equation}
where ${\kappa}^2 \equiv 8\pi/{M_{\mathrm{Pl}}}^2$, the Planck mass
being $M_{\mathrm{Pl}}  = 1.2 \times 10^{19}$ GeV. We use the
signature $(-,+,+,+)$, $g$ is the determinant of the metric tensor
$g_{\mu\nu}$,  $\Lambda$  is the
cosmological constant, $f$  is a differentiable function,  and $\mathcal{L}_\mathrm{matter}$ is the matter
Lagrangian. Note that the modified gravity action (\ref{nl1}) does not include
a new dimensional parameter. This nonlocal model has a local
scalar-tensor formulation. Introducing two scalar fields, $\eta$ and $\xi$, we can
rewrite action~(\ref{nl1}) in the following local form:
\begin{equation}
\label{anl2} S = \int d^4 x \sqrt{-g}\left\{
\frac{1}{2\kappa^2}\left[R\left(1 + f(\eta)-\xi\right) + \xi\Box\eta  -
2 \Lambda \right] + \mathcal{L}_\mathrm{matter}  \right\} \, .
\end{equation}
By varying the action (\ref{anl2}) over $\xi$, we get $\Box\eta=R$.
Substituting $\eta=\Box^{-1}R$ into action~(\ref{anl2}), one reobtains
action~(\ref{nl1}). Varying action~(\ref{anl2}) with respect to the
metric tensor $g_{\mu\nu}$, one gets
\begin{equation}
\label{nl4}
\begin{split}
&\frac{1}{2}g_{\mu\nu} \left[R\left(1 + f(\eta) -
 \xi\right)
 - \partial_\rho \xi \partial^\rho \eta - 2 \Lambda \right]
 - R_{\mu\nu}\left(1 + f(\eta) - \xi\right)+\\ &+ \frac{1}{2}\left(\partial_\mu \xi \partial_\nu \eta
+ \partial_\mu \eta \partial_\nu \xi \right)
 -\left(g_{\mu\nu}\Box - \nabla_\mu \partial_\nu\right)\left( f(\eta) -
\xi\right) + \kappa^2T_{\mathrm{matter}\, \mu\nu}=0\, ,
\end{split}
\end{equation}
where $\nabla_\mu$ is the covariant derivative and
$T_{\mathrm{matter}\,\mu\nu}$ the energy--momentum tensor of matter.

Variation of  action~(\ref{anl2}) with respect to $\eta$ yields
$\Box\xi+ f'(\eta) R=0$, where the prime denotes derivative with
respect to $\eta$.
If the scalar fields $\eta$ and $\xi$ depend on time only, then in
 the spatially flat Friedmann--Lema\^{i}tre--Robertson--Walker
metric with the interval
\begin{equation}
\label{mFr} ds^2={}-dt^2+a^2(t)\left(dx_1^2+dx_2^2+dx_3^2\right),
\end{equation}
equations~(\ref{nl4}) are equivalent to the following ones:
\begin{equation}
\label{equ1} {}- 3 H^2\left(1 + \Psi\right) + \frac{1}{2}\dot\xi
\dot\eta
 - 3H\dot\Psi + \Lambda + \kappa^2 \rho_{\mathrm{m}}=0\, ,
\end{equation}
\begin{equation}
\label{equ2} \left(2\dot H + 3H^2\right) \left(1 + \Psi\right) +
\frac{1}{2}\dot\xi \dot\eta + \ddot\Psi + 2H\dot\Psi - \Lambda +
\kappa^2 P_{\mathrm{m}}=0\, ,
\end{equation}
where $\Psi(t)=f(\eta(t))-\xi(t)$, $H=\dot{a}/a$ is the Hubble
parameter, differentiation with respect to time $t$ is denoted by a dot. For a perfect matter
fluid, we have $T_{\mathrm{matter}\, 0 0} = \rho_{\mathrm{m}}(t)$ and
$T_{\mathrm{matter}\, i j} = P_{\mathrm{m}}(t) g_{i j}$. The  equation of
state (EoS) is
\begin{equation}
\label{equ_rho} \dot\rho_{\mathrm{m}}={}-
3H(P_{\mathrm{m}}+\rho_{\mathrm{m}}).
\end{equation}

The equations of motion for the scalar fields $\eta$ and $\xi$ are as
follows
\begin{equation}
\label{equ3} \ddot \eta + 3H \dot \eta ={}- 6 \left(\dot H + 2
H^2\right) ,
\end{equation}
\begin{equation}
\label{equ4} \ddot \xi + 3H \dot \xi =6\left( \dot H + 2
H^2\right)f'(\eta) \, .
\end{equation}

Note that the considered system of equations
does not include the function $\eta$, but only $f(\eta)$, $f'(\eta)$
and time derivatives of $\eta$. Also, one can add a
constant to $f(\eta)$ and the same constant to $\xi$, without changing
of equations. So, $f(\eta)$ can be determined up to a constant.

For the model with action (\ref{anl2}), contained a perfect fluid with
a constant state parameter $w_{\mathrm{m}}$, a reconstruction procedure
has been made~\cite{Koivisto} in terms of functions of the scale
factor~$a$. For the model, describing by the initial nonlocal action
(\ref{nl1}), a technique for choosing the distortion function so as to
fit an arbitrary expansion history has been derived in~\cite{Woodard}.

Our goal  is to demonstrate how one can reconstruct $f(\eta)$ and get a
model with  exact solutions for the given Hubble parameter $H(t)$ and
the state parameter
$w_{\mathrm{m}}(t)=P_{\mathrm{m}}(t)/\rho_{\mathrm{m}}(t)$. We show
that to do this it is enough to solve only linear differential
equations.

The algorithm is as follows:
\begin{itemize}
\item Assume the explicit form of the functions $H(t)$ and
$w_{\mathrm{m}}(t)$.
\item Solve (\ref{equ_rho}) and get
$\rho_{\mathrm{m}}(t)$.
\item Solve (\ref{equ3}) and get $\eta(t)$.
\item Subtracting equation~(\ref{equ1}) from equation~(\ref{equ2}), get
a linear differential equation
\begin{equation}
\label{equPsi} \ddot\Psi+5H\dot\Psi+\left(2\dot
H+6H^2\right)(1+\Psi)-2\Lambda+\kappa^2(w_{\mathrm{m}}-1)\rho_{\mathrm{m}}=0.
\end{equation}

\item Using the known $H(t)$, $w_{\mathrm{m}}(t)$, and
$\rho_{\mathrm{m}}(t)$, solve (\ref{equPsi}) and get $\Psi(t)$.
\item
Substituting $\xi(t)=f(\eta(t))-\Psi(t)$ into equation~(\ref{equ4}), we
get a linear differential equation for $f(\eta)$
\begin{equation}
\label{equaf} f''(\eta)\dot\eta^2-12\left(\dot H + 2 H^2\right)f'(\eta)=\ddot\Psi+3H\dot\Psi.
\end{equation}
To get (\ref{equaf}) we also use the inverse function $t(\eta)$. Note
that equation~(\ref{equaf}) is a necessary condition that the model has
 solutions in the given form.

\item Solve (\ref{equaf}) and get the sought-for function $f(\eta)$.

\item Substitute the obtained function $f(\eta)$ to equation~(\ref{equ1}) and
equation~(\ref{equ2}) to check the existence of the solutions in the given
form.
\end{itemize}

\section{Nonlocal models with de Sitter solutions}

To demonstrate how the algorithm work we seek a such $f(\eta)$ that the
model has a de Sitter solution, in other words, the Hubble parameter
is a nonzero constant: $H = H_0$. In this case, equation~(\ref{equ3}) has the
following general solution:
\begin{equation}
\label{eta} \eta(t) = {}-4H_0(t-t_0) - \eta_0e^{-3H_0(t-t_0)},
\end{equation}
with integration constants $t_0$ and $\eta_0$. All equations are
homogeneous. If a  solution exists at $t_0=0$, then it
exists at an arbitrary $t_0$. So, without loss of generality we can set~$t_0=0$.

Note that equation~(\ref{equaf}) has been obtained without any restrictions
on solutions and the perfect matter fluid. To demonstrate how one can
get $f(\eta)$, which admits the existence of de Sitter solutions, in
the explicit form, we restrict ourself to the case $\eta_0=0$. In this
case, equation (\ref{equaf}) has the following form:
\begin{equation}
\label{equ7} 16H_0^2f''(\eta)-24H_0^2f'(\eta)=\Phi(\eta),
\end{equation}
where $\Phi(\eta)=\Phi(-4H_0t)\equiv \ddot\Psi+3H_0\dot\Psi$. We get
the following solution
\begin{equation}
f(\eta) =
\frac{1}{16H_0^2}\int\limits^\eta\left\{\int\limits^\zeta\Phi(\tilde{\zeta})e^{-3\tilde{\zeta}/2}d\tilde{\zeta}
+16C_3H_0^2\right\}e^{3\zeta/2}d\zeta+C_4,
\end{equation}
where $C_3$ and $C_4$ are arbitrary constants. We can fix $C_4$ without
loss of generality.

Following~\cite{1104.2692}, we consider the matter with the state
parameter $w_{\mathrm{m}}\equiv P_{\mathrm{m}}/\rho_{\mathrm{m}}$ to be a
constant, not equal to $-1$. Thus, equation~(\ref{equ_rho}) has the
following general solution
\begin{equation}
\rho_{\mathrm{m}}=\rho_0\,e^{{}-3(1+w_{\mathrm{m}})H_0t},
\end{equation}
where $\rho_0$ is an arbitrary constant. Equation~(\ref{equPsi}) has the following general solution:

 At $\, w_{\mathrm{m}}\neq 0$ and
$w_{\mathrm{m}}\neq -1/3$,
\begin{equation*}
\Psi_1(t)=C_1e^{-3H_0t}+C_2e^{-2H_0t}-1+\frac{\Lambda}{3H_0^2}-\frac{\kappa^2\rho_0(w_{\mathrm{m}}-1)}
{3H_0^2w_{\mathrm{m}}(1+3w_{\mathrm{m}})}e^{-3H_0(w_{\mathrm{m}}+1)t}\,,
\end{equation*}
\begin{equation*}
\mbox{At \ } w_{\mathrm{m}}={}-\frac{1}{3},\qquad\Psi_2(t)=C_1e^{-3H_0t}+C_2e^{-2H_0t}-1+\frac{\Lambda}{3H_0^2}
+\frac{4\kappa^2\rho_0}{3H_0}e^{-2H_0t}t\,,\qquad \quad
\end{equation*}
\begin{equation*}
\mbox{At \ } w_{\mathrm{m}}=0, \qquad \quad
\Psi_3(t)=C_1e^{-3H_0t}+C_2e^{-2H_0t}-1+\frac{\Lambda}{3H_0^2}-\frac{\kappa^2\rho_0}{H_0}e^{-3H_0t}t,\quad
\qquad \quad
\end{equation*}
where $C_1$ and $C_2$ are arbitrary constants.

Substituting the explicit form of $\Psi(t)$, we get
\begin{equation}
\label{f4}
 f_1(\eta)
=\frac{C_2}{4}e^{\eta/2}+C_3e^{3\eta/2}+C_4-\frac{\kappa^2\rho_0}{3(1+3w_{\mathrm{m}})H_0^2}
e^{3(w_{\mathrm{m}}+1)\eta/4}\,,\quad \mbox{at}\quad w_{\mathrm{m}}\neq {}-\frac{1}{3}\,,
\end{equation}
\begin{equation}
\label{fw13} \tilde{f}_1(\eta)
=\frac{C_2}{4}e^{\eta/2}+C_3e^{3\eta/2}+C_4+\frac{\kappa^2\rho_0}{4H_0^2}\left(1-\frac{1}{3}\eta\right)
e^{\eta/2}, \quad \mbox{at}\quad w_{\mathrm{m}}= {}-\frac{1}{3}\,,
\end{equation}
where $C_3$ and $C_4$ are arbitrary constants. Note that $C_2$ is an
arbitrary constant as well.

One can see that the key ingredient of all functions $f_i(\eta)$ is an
exponent function. For the models with $f(\eta)$ equal to an exponential function or a sum of exponential functions,
particular de Sitter solutions have been found
in~\cite{1104.2692,Odintsov0708}. De Sitter solutions in the case of the exponential function $f$ have
been generalized and those stability have been analysed in~\cite{EPV2011}.

\section{Conclusion}

Exact solutions play an important role in modern cosmological models,
in particular, in nonlocal cosmological
models~\cite{1104.2692,Odintsov0708,EPV2011,Biswas,AJV0711,ZS,BCK}. The
main result of this paper is the algorithm, using which one can
reconstruct $f(\Box^{-1} R)$, corresponding to the given Hubble
parameter and the state parameter of the matter. We have found that the
function $f$ corresponding to de Sitter solutions is an exponential
function or a sum of exponential functions\footnote{If the model
includes the perfect fluid with  $w_{\mathrm{m}}=-1/3$ the form of $f$
is more complicated (formula~(\ref{fw13})).}. In the case of the
exponential function $f$, expanding universe solutions $a\sim t^n$
have been found in~\cite{Odintsov0708,ZS}. We plan to analyse possible
forms of the corresponding function $f$ in future investigations.

The author is grateful to the organizers of the Dubna International
Workshop "Supersymmetries and Quantum Symmetries" (SQS'2011) for
hospitality and financial support. The author wishes to express his
thanks to Emilio~Elizalde, Sergei~D.~Odintsov, Ekaterina~O.~Pozdeeva,
Richard~P.~Woodard, and Ying-li~Zhang for useful and stimulating
discussions. The work~is supported in part by the RFBR grant
11-01-00894, by the Russian Ministry of Education and Science under
grants NSh-4142.2010.2 and NSh-3920.2012.2, and by contract CPAN10-PD12
(ICE, Barcelona, Spain).

\end{document}